\documentclass[twocolumn,conference]{IEEEtran}
\usepackage[T1]{fontenc}
\usepackage[latin9]{inputenc}
\usepackage{amsmath}
\usepackage{amsthm}
\usepackage{amssymb}
\usepackage{graphicx}

\makeatletter
\theoremstyle{plain}
\newtheorem{thm}{\protect\theoremname}
\theoremstyle{plain}
\newtheorem{lem}[thm]{\protect\lemmaname}
\theoremstyle{plain}
\newtheorem{prop}[thm]{\protect\propositionname}

\usepackage[caption=false,font=footnotesize]{subfig}

\newtheorem*{ass*}{Assumption}

\newtheorem*{apr*}{Approximation}

\makeatother

\providecommand{\lemmaname}{Lemma}
\providecommand{\propositionname}{Proposition}
\providecommand{\theoremname}{Theorem}

\begin{document}

\title{Performance Limits of Compressive Sensing Channel Estimation in Dense
Cloud RAN}

\author{\IEEEauthorblockN{Stelios~Stefanatos and Gerhard Wunder}\IEEEauthorblockA{Heisenberg Communications and Information Theory Group,\\
Freie Universität Berlin,\\
Takustr. 9, D\textendash 14195 Berlin, Germany,\\
Email: \{stelios.stefanatos, g.wunder\}@fu-berlin.de}}
\maketitle
\begin{abstract}
Towards reducing the training signaling overhead in large scale and
dense cloud radio access networks (CRAN), various approaches have
been proposed based on the channel sparsification assumption, namely,
only a small subset of the deployed remote radio heads (RRHs) are
of significance to any user in the system. Motivated by the potential
of compressive sensing (CS) techniques in this setting, this paper
provides a rigorous description of the performance limits of many
practical CS algorithms by considering the performance of the, so
called, oracle estimator, which knows \emph{a priori} which RRHs are
of significance but not their corresponding channel values. By using
tools from stochastic geometry, a closed form analytical expression
of the oracle estimator performance is obtained, averaged over distribution
of RRH positions and channel statistics. Apart from a bound on practical
CS algorithms, the analysis provides important design insights, e.g.,
on how the training sequence length affects performance, and identifies
the operational conditions where the channel sparsification assumption
is valid. It is shown that the latter is true only in operational
conditions with sufficiently large path loss exponents.
\end{abstract}

\IEEEpeerreviewmaketitle{}

\section{Introduction}

Cloud radio access network (CRAN) is considered as one of the enablers
of future cellular networks \cite{5G Wunder,C-RAN survey=000026tutorials}.
In CRAN, multiple low-complexity, low-cost remote radio heads (RRHs)
are distributed over the network coverage area and are all connected
to a central, cloud-based baseband unit whose task is to jointly process
the signals received from or send to the user equipment (UEs) in the
system. This centralized-by-design network architecture is, in principle,
able to realize the vision of large-scale, multi-cell cooperative
networks \cite{Gesbert network MIMO}. However, a major challenge
towards this goal is the need for \emph{global} channel state information
(CSI) \cite{C-RAN survey=000026tutorials,Gesbert network MIMO}, i.e.,
estimation of the quality of all channel links among the RRHs and
the UEs. With a large number of RRHs, the standard training procedure
based on orthogonal pilot sequences results in an unacceptably large
overhead \cite{LEtaief ICC}. 

Recently, various research efforts have been made towards reducing
the training overhead in CRAN. All these works are based on the premise
that, in a large-scale CRAN deployment, only a small subset of the
RRHs will have significant contribution to the downlink received energy
by any UE (similar consideration holds for the unlink as well). Therefore,
an (artificial) \emph{channel sparsification} \cite{Wunder IEEE access,channel sparsification}
is assumed at the UE side, which considers only the strongest RRH-to-UE
links for channel estimation purposes and ignores the remaining links.
This approach effectively allows for reduced training overhead using
small length pilot sequences. In \cite{LEtaief ICC}, the problem
of identifying the subset of strongest RRH channels is investigated,
assuming \emph{a priori} knowledge of large-scale fading (i.e., RRH-to-UE
distances) without, however, providing an explicit channel estimation
scheme. RRH-to-UE distances are also assumed known in \cite{locally orthoganal graph theory}
where a locally orthogonal training scheme is proposed, with short-length
training sequences obtained as the solution of a graph-based problem,
which, however, depends on the positions of RRHs and UEs. Another
approach is considered in \cite{Lau ICC,Quek ICC}, where CSI is treated
as a sparse vector and estimation algorithms inspired by the compressive
sensing (CS) framework \cite{CS textbook} are applied. Good estimation
performance with small length training sequences is demonstrated only
via simulations and for certain system setups. All these works provide
valuable insights on the effectiveness of small-length training sequences,
however, the \emph{operational conditions, }e.g., propagation losses
and small-scale fading statistics, under which the channel sparsification
assumption is applicable are not clear. 

In this paper, the performance potential of applying CS techniques
towards reducing downlink training overhead in large-scale CRAN deployments
is investigated. In particular, for a CS-motivated training sequence
design that is independent of number and positions of RRHs and UEs,
the performance of the, so called, oracle estimator is analytically
obtained. The oracle estimator has \emph{a priori} knowledge of the
set of RRHs with the strongest links, but not their channel values,
and its performance serves as a optimistic estimate for the performance
of many practical CS estimation algorithms (that have no \emph{a priori}
information) \cite{oracle_based_sensing_matrix_optimization,Coluccia}. 

By employing tools from stochastic geometry, a closed-form upper bound
of the oracle estimator mean squared error performance is obtained,
that is averaged over the distribution of RRH positions and channel
statistics, and is \emph{independent} of the UE position within the
CRAN deployment. The bound expression is tight for large scale, dense
CRAN deployments and provides insights on how estimation performance
is affected by (a) system \emph{design }parameters, e.g., training
sequences length and number of (strongest) channels to be estimated\emph{,
}and (b)\emph{ operational} \emph{conditions}, e.g., RRH density and
channel statistics. One of the main takeaways of the analysis is that
the channel sparsification assumption is not valid for operational
conditions with small path loss exponent, suggesting that training
overhead reduction is not possible in these cases, irrespective of
the approach to do so (not necessarily CS-based).

\emph{Notation}: $[N]$ denotes the set $\{1,2,\ldots,N\}$. $(\cdot)^{T},(\cdot)^{H}$
denote transpose and Hermitian transpose, respectively. $|\cdot|$
will be used to denote the modulus of a complex number, or the (Lebesgue)
measure of a set, depending on context. The Frobenius norm of a vector
(possibly infinite-dimensional) will be denoted as $\|\cdot\|$. The
$n$-th element of an $N$-dimensional vector $\mathbf{x}$ is denoted
as $[\mathbf{x}]_{n},n\in[N]$, and $\mathbf{I}$ is the identity
matrix of appropriate dimensions. $\mathbf{X}_{\mathcal{B}}$ ($\mathbf{x}_{\mathcal{B}}$)
is the matrix (vector) obtained by considering only the columns (elements)
indicated by the set $\mathcal{B}$. 

\section{System Model and Performance Metrics}

A dense CRAN deployment is considered employing $N_{\text{RRH}}$
single-antenna RRHs whose positions are independently and uniformly
distributed over a large deployment area $\mathcal{A}\subset\mathbb{R}^{2}$,
resulting in an RRH \emph{density} (i.e., average number of RRHs per
unit area of deployment) equal to $\lambda\triangleq N_{\text{RRH}}/|\mathcal{A}|$.
Let $\Phi\subset\mathbb{R}^{2}$ denote the set of RRH positions and
$h_{x}\in\mathbb{C},$ denote the baseband, flat-fading channel on
a single OFDM subcarrier of the link between an RRH positioned at
$x\in\Phi$ and a UE located at an arbitrary position $x_{u}\in\mathbb{R}^{2}$.
This channel is commonly modeled as \cite{channel sparsification,locally orthoganal graph theory,Haenggi Ganti book}
\begin{equation}
h_{x}=c_{x}\|x-x_{u}\|^{-\alpha/2},x\in\Phi,\label{eq:channel model}
\end{equation}

\noindent where $\alpha>2$ is the (deterministic) \emph{path loss
exponent}, and $c_{x}\in\mathbb{C}$ represents small and/or large
scale fading effects. The variables $\{c_{x}\}_{x\in\Phi}$ are assumed
to be independent, identically distributed (i.i.d.) with $\mathbb{E}(c_{x})=0$
and $\mathbb{E}(|c_{x}|^{4/\alpha})<\infty$. The seemingly random
last assumption is only a technical one that will be employed later
in the analysis and is actually satisfied by many models used in practice
for the statistics of $c_{x}$ such as Rayleigh and lognormal fading. 

Towards acquiring downlink CSI, each RRH transmits a signature training
(pilot) sequence that is known by each UE in the system. It is assumed
that the transmission of training sequences is done synchronously
among RRHs and lasts for $N_{p}$ symbols. Let $\mathbf{p}_{x}\in\mathbb{C}^{N_{p}}$
denote the column vector of $N_{p}$ pilot symbols transmitted by
the RRH positioned at $x\in\Phi$ with a normalized average transmit
power equal to $\frac{1}{N_{p}}\mathbb{E}(\mathbf{p}_{x}^{H}\mathbf{p}_{x})=1$,
for all $x\in\Phi$. The received signal $\mathbf{y}\in\mathbb{C}^{N_{p}}$
of the UE at $x_{u}$ during the training period equals
\begin{align}
\mathbf{y} & =\sum_{x\in\Phi}h_{x}\mathbf{p}_{x}+\mathbf{w}=\mathbf{P}\mathbf{h}+\mathbf{w}\label{eq:system model}
\end{align}

\noindent where $\mathbf{h}\in\mathbb{C}^{N_{\text{RRH}}}$ is a vector
containing the $N_{\text{RRH}}$ channel values in some arbitrary
order, $\mathbf{P}\in\mathbb{C}^{N_{p}\times N_{\text{RRH}}}$ is
the \emph{pilot matrix }whose columns are the RRH training sequences
with the same ordering as the elements of $\mathbf{h}$, and $\mathbf{w}\in\mathbb{C}^{N_{p}}$
represents noise, consisting of i.i.d. complex Gaussian elements of
zero mean and variance $\sigma_{w}^{2}$. 

Under the conventional, orthogonal training sequence approach, i.e.,
when it holds $\mathbf{P}^{H}\mathbf{P}=N_{p}\mathbf{I}$, global
CSI can be acquired at the UE by simply computing $\frac{1}{N_{p}}\mathbf{P}^{H}\mathbf{y}$.
However, the orthogonality condition requires $N_{p}\geq N_{\text{RRH}}$,
which results in an unacceptable pilot overhead when $N_{\text{RRH}}\gg1$.
This observation motivates the investigation of channel estimation
procedures under the condition $N_{p}<N_{\text{RRH}}$, i.e., with
small-length training sequences, that are \emph{necessarily} \emph{non-orthogonal}.
Even though this approach does not allow for a reliable estimate of
the global CSI (less observations than unknowns), it is motivated
by the channel sparsification assumption \cite{channel sparsification},
namely, that out of the $N_{\text{RRH}}$ elements of $\mathbf{h}$,
only $s\ll N_{\text{RRH}}$ are of significant modulus, whereas the
others can be safely considered as equal to zero. This, in turn, motivates
the use of CS techniques that achieve reliable estimates with $N_{p}=\mathcal{O}(s)$
\cite{CS textbook}.

Unfortunately, the performance of CS estimation algorithms is difficult
to describe accurately. For this reason, a commonly used approach
is to lower bound their performance by considering the performance
of the, so called, \emph{oracle estimator} \cite{Coluccia}, which
is essentially the standard least squares (LS) channel estimator of
the $s$ largest-modulus elements of $\mathbf{h}$ \emph{under the
assumption that their positions within $\mathbf{h}$ are known}. In
addition to providing a performance bound, the oracle estimator performance
will provide insights on how accurate the channel sparsification assumption
is under various operational conditions, which is critical for the
effectiveness of any short-length training approach (not necessarily
based on CS techniques).

Let $\mathcal{S}\subseteq[N_{\text{RRH}}]$ denote the set of $s\triangleq|\mathcal{S}|\geq1$
indices of largest-modulus elements in $\mathbf{h}$, which are known
by the oracle estimator. The value of $s$ is treated here as a free
design parameter to be optimized in the following. Note that it is
not clear at this point of the discussion how to optimally choose
$s$ due to the conflicting requirements of small-length training
sequences and accurate global CSI, suggesting small and large values
for $s$, respectively. Assuming that $\mathbf{P}_{\mathcal{S}}$
is full column rank (which implies $N_{p}\geq s$), the LS estimate
of $\mathbf{h}_{\mathcal{S}}$ equals 
\begin{align}
\hat{\mathbf{h}}_{\mathcal{S}} & \triangleq\left(\mathbf{P}_{\mathcal{S}}^{H}\mathbf{P}_{\mathcal{S}}\right)^{-1}\mathbf{P}_{\mathcal{S}}^{H}\mathbf{y}\nonumber \\
 & =\mathbf{h}_{\mathcal{S}}+\left(\mathbf{P}_{\mathcal{S}}^{H}\mathbf{P}_{\mathcal{S}}\right)^{-1}\mathbf{P}_{\mathcal{S}}^{H}\left(\mathbf{P}_{\bar{S}}\mathbf{h}_{\bar{\mathcal{S}}}+\mathbf{w}\right),\label{eq:LS_estimate_formula}
\end{align}
where $\bar{\mathcal{S}}\triangleq[N_{\text{RRH}}]\setminus\mathcal{S}$
denotes the indices of elements of $\mathbf{h}$ that are not estimated.
Note that since $\mathbf{P}$ consists of non-orthogonal columns,
the product $\mathbf{P}_{\mathcal{S}}^{H}\mathbf{P}_{\bar{S}}$ in
(\ref{eq:LS_estimate_formula}) does not vanish, therefore, the pilot
transmissions from RRHs corresponding to $\mathcal{\bar{S}}$ act
as \emph{interference }for the channel estimation procedure.

It directly follows from the channel statistics assumptions that the
error term of the LS estimate appearing on the right hand side (RHS)
of (\ref{eq:LS_estimate_formula}) is zero mean, i.e., the LS estimate
is unbiased, irrespective of the choice of $\mathcal{S}$ and $\mathbf{P}$.
However, the latter do affect the \emph{average} mean squared error
($\mathsf{MSE}_{\text{av}}$) of the estimate, which provides a measure
of the estimation accuracy for each element of $\hat{\mathbf{h}}_{\mathcal{S}}$
and is defined as
\begin{align*}
\mathsf{MSE}_{\text{av}} & \triangleq\frac{1}{s}\mathbb{E}\left(\|\mathbf{h}_{\mathcal{S}}-\hat{\mathbf{h}}_{\mathcal{S}}\|^{2}\right)\\
 & =\frac{1}{s}\mathbb{E}\left(\left\Vert \left(\mathbf{P}_{\mathcal{S}}^{H}\mathbf{P}_{\mathcal{S}}\right)^{-1}\mathbf{P}_{\mathcal{S}}^{H}\left(\mathbf{P}_{\bar{S}}\mathbf{h}_{\bar{\mathcal{S}}}+\mathbf{w}\right)\right\Vert ^{2}\right).
\end{align*}

Another related metric for evaluating the channel estimation performance,
of particular importance in a CRAN setting, is the \emph{total} mean
squared error ($\mathsf{MSE}_{\text{tot}}$), defined as
\begin{align}
\mathsf{MSE}_{\text{tot}} & \triangleq\mathbb{E}\left(\|\mathbf{h}_{\mathcal{\bar{S}}}\|^{2}\right)+\mathbb{E}\left(\|\mathbf{h}_{\mathcal{S}}-\hat{\mathbf{h}}_{\mathcal{S}}\|^{2}\right)\nonumber \\
 & =\mathbb{E}\left(\|\mathbf{h}_{\mathcal{\bar{S}}}\|^{2}\right)+s\mathsf{MSE}_{\text{av}}.\label{eq:MSE_tot_definition}
\end{align}

\noindent $\mathsf{MSE}_{\text{tot}}$ provides an indication of how
well the LS estimate $\hat{\mathbf{h}}_{\mathcal{S}}$ (of $s$ elements)
can be used to approximate the \emph{complete} channel vector $\mathbf{h}$
(of $N_{\text{RRH}}$ elements) under the channel sparsification assumption,
i.e., by treating the elements of $\mathbf{h}_{\bar{\mathcal{S}}}$
as zeros. 

\section{Mean Squared Error Performance of the Oracle Estimator}

The $\mathsf{MSE}$ performance of the oracle channel estimate depends
on multiple parameters, namely, 
\begin{itemize}
\item positions $\Phi$ of the RHHs; 
\item position $x_{u}$ of the UE in consideration; 
\item path loss exponent $\alpha$;
\item channel fading statistics; 
\item number $s$ of estimated channels;
\item pilot matrix $\mathbf{P}$.
\end{itemize}
From this set of parameters, only the last two are \emph{design} parameters
that should, in principle, be optimized given the values of the remaining
parameters describing \emph{operational conditions}. However, such
a design approach is of little practical interest as it is dependent
on the UE position, whereas practical systems employ a fixed set of
training sequences that (hopefully) results in a good channel estimate
at the UE side irrespective of its position.

Towards\emph{ }such a training sequence design and analysis, the following
assumption on the pilot matrix will be employed.

\begin{ass*} 
The elements of the pilot matrix $\mathbf{P}$ are generated as i.i.d. real-valued Gaussian variables of zero mean and variance $1$.
\end{ass*}

Although this design choice is not necessarily optimal, it is motivated
by noting that, in the observation model of (\ref{eq:system model}),
$\mathbf{P}$ acts as a \emph{sensing} matrix in a standard CS estimation
setting. It is well known that a sensing matrix generated as per the
previous assumption is a good design choice \cite{CS textbook}. Note
that $N_{p}$ is not specified in the above assumption, i.e., it is
a free design parameter to be optimized in the following.

For a Gaussian pilot matrix, a preliminary characterization of the
$\mathsf{MSE}$, averaged over all possible pilot matrix realizations
can be obtained.
\begin{lem}
\label{lem:basic oracle MSE}With Gaussian training sequences, the
average mean squared error performance of the oracle estimator considering
the $s\leq N_{p}-4$ largest-modulus elements of $\mathbf{h}$ equals
\begin{equation}
\mathsf{MSE}_{\text{\emph{av}}}=\frac{\mathbb{E}\left(\|\mathbf{h}_{\bar{\mathcal{S}}}\|^{2}\right)+\sigma_{w}^{2}}{N_{p}-s-1},\label{eq:MSE white noise}
\end{equation}

\noindent whereas the total mean squared error equals
\begin{equation}
\mathsf{MSE}_{\text{\emph{tot}}}=\frac{(N_{p}-1)\mathbb{E}\left(\|\mathbf{h}_{\bar{\mathcal{S}}}\|^{2}\right)+\sigma_{w}^{2}}{N_{p}-s-1}.\label{eq:MSE_tot white noise}
\end{equation}
\end{lem}
\begin{IEEEproof}
Direct application of \cite[Theorem 1]{Coluccia} for the LS estimate
of $\mathbf{h}_{\mathcal{S}}$ based on the observation $\mathbf{y}=\mathbf{P}_{\mathcal{S}}\mathbf{h}_{\mathcal{S}}+(\mathbf{P}_{\bar{S}}\mathbf{h}_{\bar{\mathcal{S}}}+\mathbf{w})$
yields
\begin{align*}
\mathsf{MSE}_{\text{av}} & =\frac{\mathbb{E}\left(\|\mathbf{P}_{\bar{S}}\mathbf{h}_{\bar{\mathcal{S}}}+\mathbf{w}\|^{2}\right)}{N_{p}\left(N_{p}-s-1\right)}=\frac{\mathbb{E}\left(\|\mathbf{P}_{\bar{S}}\mathbf{h}_{\bar{\mathcal{S}}}\|^{2}\right)+N_{p}\sigma_{w}^{2}}{N_{p}\left(N_{p}-s-1\right)}.
\end{align*}
 Noting that $\mathbb{E}\left(\|\mathbf{P}_{\bar{\mathcal{S}}}\mathbf{h}_{\bar{\mathcal{S}}}\|^{2}\right)=\mathbb{E}\left(\mathbf{h}_{\bar{\mathcal{S}}}^{H}\mathbb{E}\left(\mathbf{P}_{\bar{\mathcal{S}}}^{H}\mathbf{P}_{\bar{\mathcal{S}}}\right)\mathbf{h}_{\bar{\mathcal{S}}}\right)=N_{p}\mathbb{E}\left(\|\mathbf{h}_{\bar{\mathcal{S}}}\|^{2}\right)$,
since $\mathbb{E}\left(\mathbf{P}_{\bar{\mathcal{S}}}^{H}\mathbf{P}_{\bar{\mathcal{S}}}\right)=N_{p}\mathbf{I}$
by construction, leads to (\ref{eq:MSE white noise}). Substituting
(\ref{eq:MSE white noise}) into (\ref{eq:MSE_tot_definition}) leads
to (\ref{eq:MSE_tot white noise}).
\end{IEEEproof}
\emph{Remark}: The upper bound $N_{p}-4$ for $s$ appearing in Lemma
\ref{lem:basic oracle MSE} is a technical one \cite{Coluccia} and
has no effect on the design, as it will be shown that values of $s$
far smaller than this bound are of interest.

The expectation appearing on the RHS of (\ref{eq:MSE white noise})
is over the statistics of $\mathbf{h}_{\bar{\mathcal{S}}}$ for a
given UE position $x_{u}\in\mathbb{R}^{2}$. In order to obtain a
robust (i.e., worst case) design that is independent of the UE position,
it is natural to consider the case where $x_{u}$ lies ``in the middle''
of the RRH deployment area since, in that case, the effect of the
interference is expected to be greater, compared to the case when
the UE lies on the edge of the RRH deployment area. However, it is
not clear which is this ``middle'' position and how it depends on
the RRH deployment area $\mathcal{A}$. In order to avoid these issues,
the following approximation on the distribution of RRH positions will
be considered.

\begin{apr*} 
The positions of the RRHs are distributed over all $\mathbb{R}^2$ as a homogeneous Poisson point process (HPPP) $\tilde{\Phi} \subset \mathbb{R}^2$ of density $\lambda$.
\end{apr*}

This approximation can be viewed as extending the original RRH deployment
area $\mathcal{A}$ to the whole $\mathbb{R}^{2}$, thus implying
an infinite number of RRHs (instead of $N_{\text{RRH}}$), however,
with the same density as the actual deployment. Clearly, this approximation
leads to a pessimistic performance analysis, since, for a given $\mathcal{S}$,
the power of the interference term in (\ref{eq:LS_estimate_formula})
statistically increases for any UE position. However, this approximation
is expected to be accurate under sufficiently large $N_{\text{RRH}}$
and/or $\alpha$ for a UE lying ``in the middle'' of the actual
deployment. Note that, due to the stationarity of the HPPP \cite{Haenggi Ganti book},
the statistics of $\mathbf{h}$ (over fading and RRH positions) are
independent of the UE position, hence, it will be assumed for convenience
that $x_{u}=(0,0)$, i.e, the UE in consideration lies at the origin
of the plane.

Under the HPPP approximation, a closed-form expression for the term
$\mathbb{E}(\|\mathbf{h}_{\bar{\mathcal{S}}}\|^{2})$ appearing in
(\ref{eq:MSE white noise}) and (\ref{eq:MSE_tot white noise}) can
be obtained.
\begin{prop}
\label{thm: covariance of interference}Under the HPPP approximation
for the RRH distribution, it holds
\begin{equation}
\mathbb{E}(\|\mathbf{h}_{\bar{\mathcal{S}}}\|^{2})=\frac{2\left(\pi\lambda\mathbb{E}\left(|c_{x}|^{4/\alpha}\right)\right)^{\frac{\alpha}{2}}\Gamma\left(s+1-\frac{\alpha}{2}\right)}{(s-1)!\left(\alpha-2\right)},\label{eq:covariance of interference}
\end{equation}
for all $s>\frac{\alpha}{2}-1$, where $\Gamma(t)\triangleq\int_{0}^{\infty}z^{t-1}e^{-z}dz$
is the Gamma function.
\end{prop}
\begin{IEEEproof}
See Appendix A.
\end{IEEEproof}
\emph{Remark}: The restriction $s>\frac{\alpha}{2}-1$appearing in
Prop. \ref{thm: covariance of interference} is only a minor one since,
as will be shown in the following, much greater values are of interest.

By combining Lemma \ref{lem:basic oracle MSE} with Prop. \ref{thm: covariance of interference},
a closed form upper bound for the oracle estimator $\mathsf{MSE}$
directly follows.
\begin{prop}
\label{prop: Oracle_MSE_final_expression}Consider a CRAN deployment
with an arbitrary number of RRHs independently and uniformly distributed
over a large deployment area employing Gaussian training sequences
of length $N_{p}$. For any UE in the system, the $\mathsf{MSE}_{\text{\emph{av}}}$
and $\mathsf{MSE}_{\text{\emph{tot}}}$ of the oracle estimate considering
the $s\in(\frac{\alpha}{2}-1,N_{p}-4]$ largest-modulus elements of
$\mathbf{h}$ is upper bounded as
\begin{equation}
\mathsf{MSE}<\beta\frac{2\left(\pi\lambda\mathbb{E}\left(|c_{x}|^{4/\alpha}\right)\right)^{\frac{\alpha}{2}}\Gamma\left(s+1-\frac{\alpha}{2}\right)}{(N_{p}-s-1)(\alpha-2)(s-1)!}+\gamma\frac{\sigma_{w}^{2}}{N_{p}-s-1},\label{eq:MSE_bound}
\end{equation}

\noindent with $\{\beta=1,\gamma=1\}$ and $\{\beta=N_{p}-1,\gamma=s\}$,
for $\mathsf{MSE}_{\text{\emph{av}}}$ and $\mathsf{MSE}_{\text{\emph{tot}}}$,
respectively.
\end{prop}
The following remarks regarding the upper bound of the $\mathsf{MSE}$
are in order:
\begin{itemize}
\item It is independent of the number $N_{\text{RHH}}$ of actually deployed
RRHs. This is due to the infinite number of RHHs implied by the HPPP
approximation.
\item Its dependence on the channel fading statistics is only via the moment
$\mathbb{E}\left(|c_{x}|^{4/\alpha}\right)$, i.e., fading statistics
with the same moment are indistinguishable in terms of estimation
performance.
\item For $\alpha\rightarrow2$, it grows unbounded, irrespective of the
values of the other parameters, since the sum power of the interfering
signals becomes very large. 
\item When the noise effect is negligible (i.e., $\sigma_{w}^{2}=0$), it
grows as $\mathcal{O}(\lambda^{\alpha/2})$, implying that RRH densification
(i.e., increasing $\lambda$) under conditions with high path loss
exponent has more impact on the estimation performance than under
small path loss exponents.
\item For $N_{p}\rightarrow\infty$ it scales as $\mathcal{O}(1/N_{p})$
and $\mathcal{O}(1)$ for $\mathsf{MSE}_{\text{av}}$ and $\mathsf{MSE}_{\text{tot}}$,
respectively, implying that, \emph{for a fixed $s$}, increasing $N_{p}$
is beneficial only in terms of accuracy of $\hat{\mathbf{h}}_{\mathcal{S}}$
but has no effect on the accuracy of the complete channel vector estimate
since, by default, the elements of $\hat{\mathbf{h}}_{\bar{\mathcal{S}}}$
are set to zero irrespective of $N_{p}$.
\item Under conventional, orthogonal training, it is easy to see that $\mathsf{MSE}_{\text{av}}=\sigma_{w}^{2}/N_{p}$,
irrespective of $s$. In contrast, the upper bound $\mathsf{MSE}$
expression for the non-orthogonal case indicates that there exists
a value of $s$ that achieves minimum $\mathsf{MSE}$ by optimally
balancing between two contradicting requirements: (a) consideration
of a small $s$, suggested by estimation theory in a multiple parameter
estimation setting in the presence of noise/interference, and (b)
consideration of a large $s$ in order to reduce both the number and
interference power of ignored RRHs.
\end{itemize}
In regards to the last remark, the closed form expression provided
by Prop. \ref{prop: Oracle_MSE_final_expression} allows for an efficient
optimization with respect to (w.r.t.) $s$, without the need to resort
in time-consuming simulations. In particular, for the case of $\sigma_{w}^{2}=0$
(i.e., no noise), a closed form asymptotic characterization of the
optimal $s$ is available.
\begin{prop}
\label{prop:asymptotic optimal values}With $\sigma_{w}^{2}=0$, the
upper bounds for $\mathsf{MSE}_{\text{av}}$ and $\mathsf{MSE}_{\text{tot}}$
of the oracle estimator are minimized for the same value of $s$,
which is asymptotically equal to (for $N_{p}\rightarrow\infty$) 
\begin{equation}
s^{*}\sim\frac{(\alpha-2)(N_{p}-1)}{\alpha},\label{eq:optimal_s_asymptotic}
\end{equation}

\noindent resulting in the asymptotic upper bound for the minimum
$\mathsf{MSE}$ 
\begin{equation}
\text{\ensuremath{\mathsf{MSE}}}^{*}<\beta\left(\frac{\alpha\pi\mathbb{E}\left(|c_{x}|^{4/\alpha}\right)}{(\alpha-2)}\cdot\frac{\lambda}{N_{p}-1}\right)^{\alpha/2},\label{eq:minimum_MSE_asymptotic}
\end{equation}

\noindent with $\beta=1$ and $\beta=N_{p}-1$, for $\mathsf{MSE}_{\text{av}}$
and $\mathsf{MSE}_{\text{tot}}$, respectively.
\end{prop}
\begin{IEEEproof}
With $\sigma_{w}^{2}=0$, it directly follows from Prop. \ref{prop: Oracle_MSE_final_expression}
that the optimal $s$ for both $\mathsf{MSE}_{\text{av}}$ and $\mathsf{MSE}_{\text{tot}}$
is the one maximizing the term $\frac{\Gamma\left(s+1-\frac{\alpha}{2}\right)}{(N_{p}-s-1)(s-1)!}$.
Numerical examination of this term indicates that the optimal $s$
is an increasing function of $N_{p}$, which, in turn, suggests that
for $N_{p}\gg1$ it will also be $s^{*}\gg1$. Noting that $\frac{\Gamma(s+1-\alpha/2)}{(s-1)!}\sim s^{1-\alpha/2}$
for $s\gg1$, $s^{*}$ can be found as the minimizer of $s^{1-\alpha/2}/(N_{p}-s-1)$,
which is given by (\ref{eq:optimal_s_asymptotic}). Substituting the
latter into the (\ref{eq:MSE_bound}) results in (\ref{eq:minimum_MSE_asymptotic}).
\end{IEEEproof}
As it will be shown in the next section, the asymptotic expressions
given in Prop. \ref{prop:asymptotic optimal values} are very accurate
even for moderate values of $N_{p}$. Interestingly, the above result
shows that when interference is the main source of estimation error,
the optimal number of estimated channels is increasing with $\alpha$
and/or $N_{p}$ and is the same for both average and total $\mathsf{MSE}$
performance metrics. However, even with optimal $s$, the $\mathsf{MSE}$
performance severely degrades when $\alpha\rightarrow2$, indicating
\emph{a fundamental performance limitation of the non-orthogonal training
approach under sufficiently small $\alpha$} as the channel sparsification
assumption in that regime is not applicable. For $\alpha>2$, the
optimal upper bound scales as $\mathcal{O}(N_{p}^{-\alpha/2})$ and
$\mathcal{O}(N_{p}^{1-\alpha/2})$ for $\mathsf{MSE}_{\text{av}}$
and $\mathsf{MSE}_{\text{tot}}$, respectively, showing that increasing
$N_{p}$ does help in improving both average \emph{and} total estimation
performance, with a more prominent effect for larger values of $\alpha$.
This suggests that \emph{for operational conditions with sufficiently
large $\alpha$, good oracle estimator performance is possible with
$N_{p}\ll N_{\text{RRH}}$}. This, in turn, implies that the channel
sparsification assumption is valid in the large $\alpha$ regime,
which can be exploited to reduce training signaling overhead by, e.g.,
use of CS estimation algorithms. 

\section{Numerical Results}

\begin{figure}[t]
\noindent \centering{}\includegraphics[width=0.78\columnwidth]{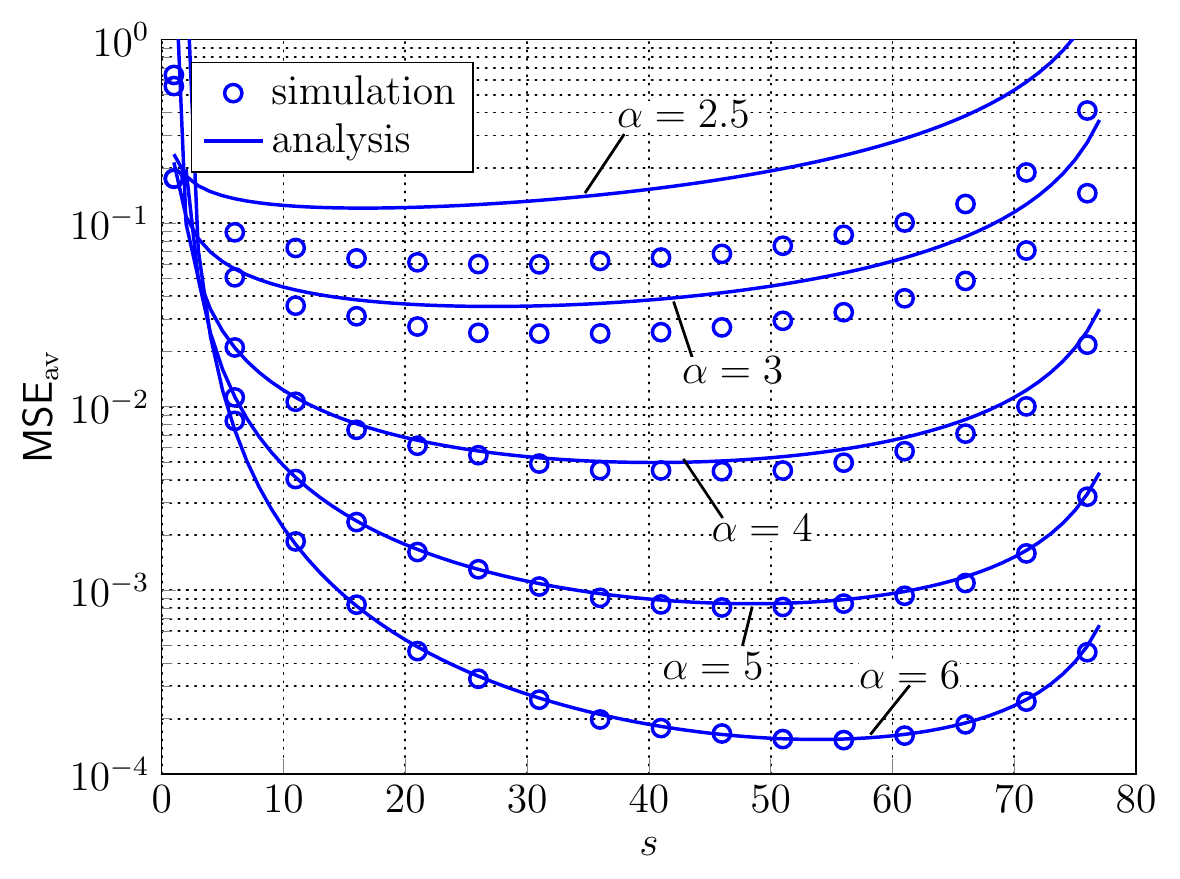}\caption{$\mathsf{MSE}_{\text{av}}$ vs $s$ (number of estimated channels)
of the oracle estimator. Solid lines corresponds to the bound of Prop.
\ref{prop: Oracle_MSE_final_expression} and markers are obtained
by simulation.\label{fig:MSE-vs-s}}
\end{figure}

\emph{1) Comparison of upper bound on $\mathsf{MSE}_{\text{av}}$
with simulated performance}: A CRAN deployment of $N_{\text{RRH}}=500$
RRHs with a density $\lambda=1$, uniformly distributed over a rectangular
area $\mathcal{A}$ was simulated. The channel fading variables $c_{x}$
where generated as i.i.d. complex Gaussians of zero mean and variance
$1$ (Rayleigh fading), and a training sequence length of $N_{p}=81$
was considered with Gaussian pilot symbols, as described in the previous
section. Note that this training sequence length corresponds to an
$83.8\%$ overhead reduction compared to the minimum requirements
under orthogonal training. Figure \ref{fig:MSE-vs-s} shows the $\mathsf{MSE}_{\text{av}}$
of the oracle estimator for a UE located at the center of $\mathcal{A}$,
as a function of the number $s$ of estimated channels, for various
values of $\alpha$ and with $\sigma_{w}^{2}=0$. The $\mathsf{MSE}_{\text{av}}$
values were obtained by averaging over multiple independent realizations
of RRH positions and channels. In addition, the upper bound of the
$\mathsf{MSE}_{\text{av}}$, given in Prop. \ref{prop: Oracle_MSE_final_expression}
(with $\mathbb{E}(|c_{x}|^{4/\alpha})=\Gamma(1+2/\alpha)$), is also
shown.

It can be seen that the $\mathsf{MSE}$ performance depends strongly
on $\alpha$, with larger values being beneficial due to reduced interference
power, even though the power of the strongest channels is also reduced
in this case as well. This provides a strong motivation for reduced
length training sequences under operational conditions with large
$\alpha$ (e.g., in urban areas). In addition, the existence of an
optimal value of $s$,  depending on $\alpha$, is clear. Note that
the analytical upper bound provides an excellent indicator of the
performance for $\alpha\geq4$ but becomes increasingly loose for
values of $\alpha\rightarrow2$. However, the bound is still able
to follow the actual performance trends rather closely.

\begin{figure}[t]
\noindent \centering{}\includegraphics[width=0.8\columnwidth]{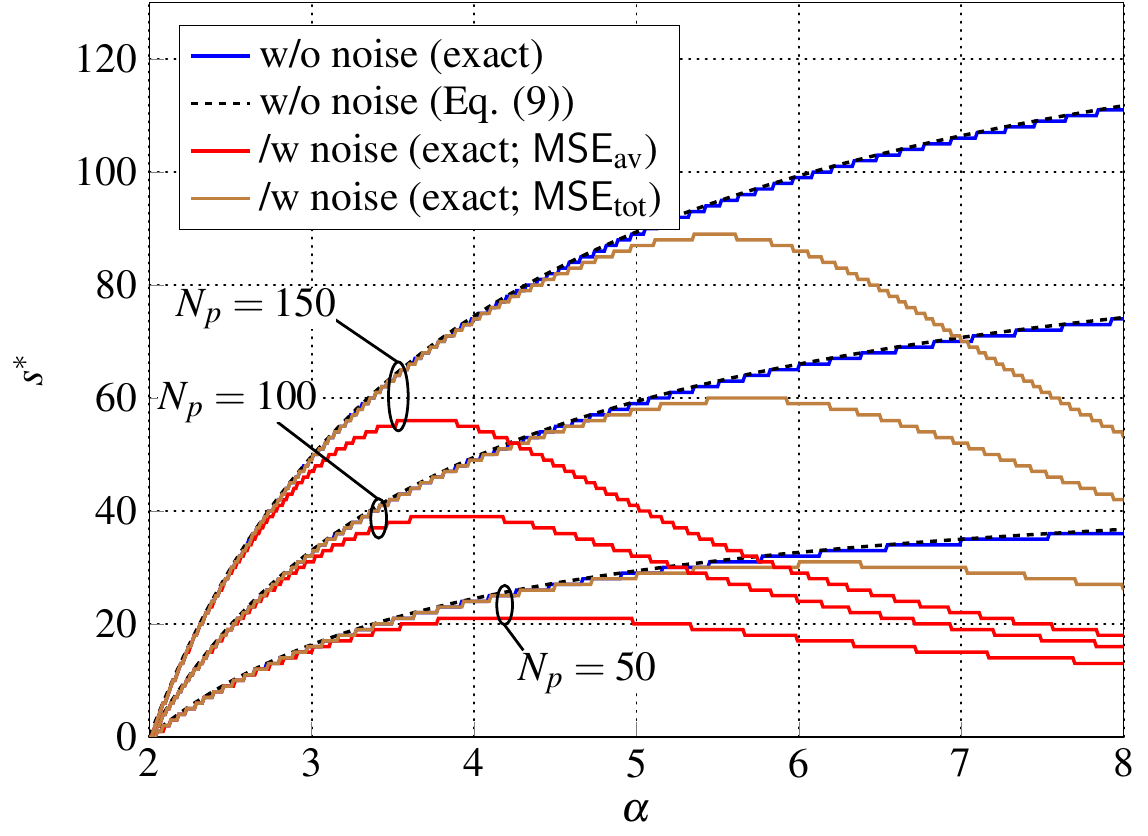}\caption{Optimal $s$ that minimizes the analytical upper bounds for the $\mathsf{MSE}_{\text{av}}$
and $\mathsf{MSE}_{\text{tot}}$ as a function of $\alpha$ with and
without noise.\label{fig:Optimal s}}
\end{figure}

\emph{2) Optimal number of estimated channels}: Figure \ref{fig:Optimal s}
shows the optimal value of $s$ ($s^{*}$) that minimizes the analytical
upper bounds for the $\mathsf{MSE}_{\text{av}}$ and $\mathsf{MSE}_{\text{tot}}$
given in Prop. \ref{prop: Oracle_MSE_final_expression}, as a function
of $\alpha$ and for various values of $N_{p}$. A case with no noise
$(\sigma_{w}^{2}=0)$ and a case with noise ($\sigma_{w}^{2}=0.1$)
is depicted. As was observed in Prop. \ref{prop:asymptotic optimal values},
the optimal $s$ is the same for both $\mathsf{MSE}_{\text{av}}$
and $\mathsf{MSE}_{\text{tot}}$ with no noise. For that case, the
asymptotic expression of (\ref{eq:optimal_s_asymptotic}) is also
depicted, which can be seen to be an excellent indicator of $s^{*}$,
even for moderate values of $N_{p}$. Interestingly, when noise is
present, the value of $s^{*}$, even though increasing with $N_{p}$,
depicts a non monotonic behavior w.r.t. $\alpha$. In particular,
for the small $\alpha$ regime, $s^{*}$ increases with $\alpha$
since the interference power is reduced allowing for the reliable
estimation of more (strong) channels. However, beyond a certain value
of $\alpha$, most of the channel powers become comparable to the
noise level, therefore, it is beneficial to concentrate the estimation
efforts on a limited set of strongest channels. 

\begin{figure}[t]
\noindent \centering{}\includegraphics[width=0.8\columnwidth]{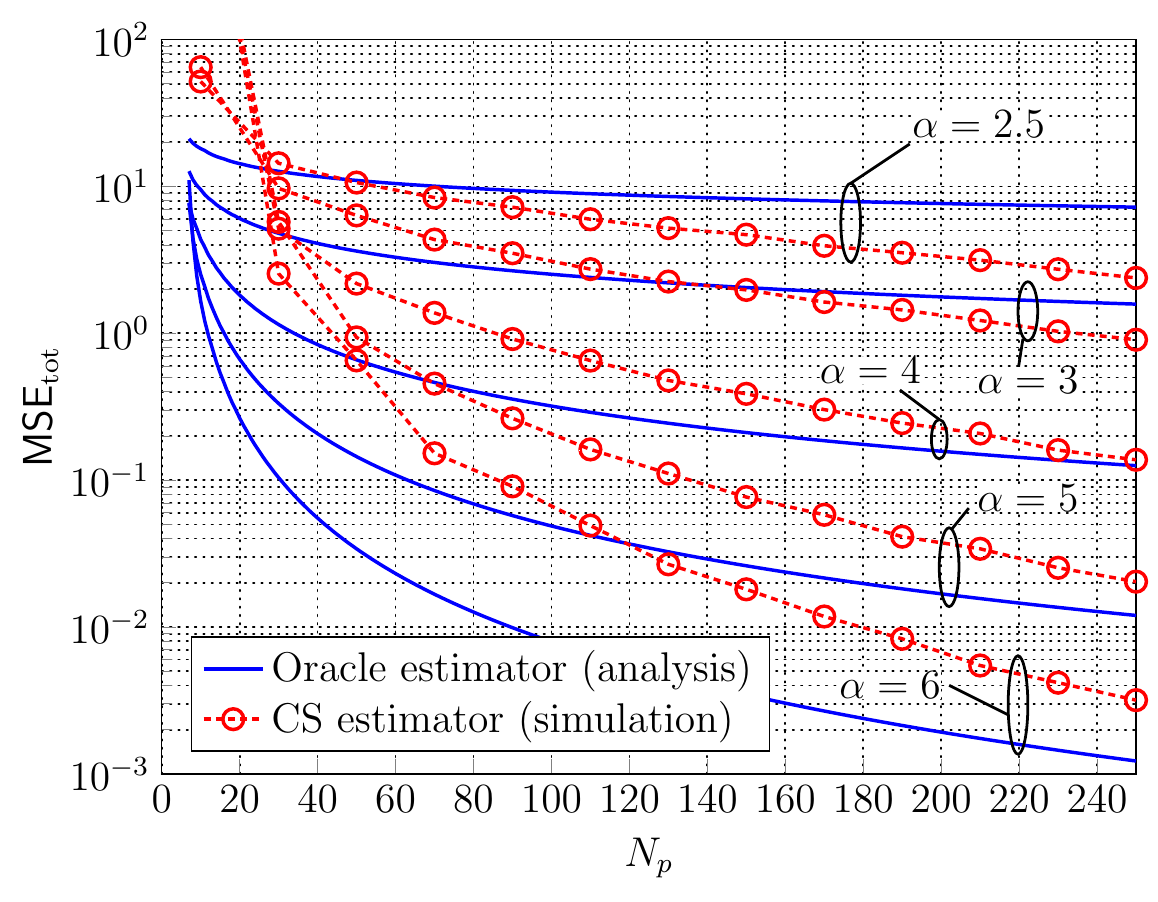}\caption{$\mathsf{MSE}_{\text{tot}}$ for the CS-based channel estimator as
a function of the training sequence length $N_{p}$. The analytical
upper bound for the $\mathsf{MSE}_{\text{tot}}$ of the oracle estimator
with optimized $s$ is also shown. \label{fig: CS estimator}}
\end{figure}

\emph{3) Compressive sensing channel estimation}: When the UE has
no \emph{a priori} knowledge of the set of largest-modulus elements
of $\mathbf{h}$, a CS estimation approach can be employed for estimating
the complete channel vector $\mathbf{h}$ as long as the channel sparsification
assumption is valid \cite{Lau ICC,Quek ICC}. Figure \ref{fig: CS estimator}
shows the simulated $\mathsf{MSE}_{\text{tot}}$ performance of the
estimator applying the standard Basis Pursuit algorithm on the received
signal $\mathbf{y}$ \cite{CS textbook}, as a function of $N_{p}$
and for various values of $\alpha$, for the exact same system setup
considered in Fig. \ref{fig:MSE-vs-s} ($N_{\text{RRH}}=500$, $\sigma_{w}^{2}=0$).
It can be seen that the CS estimator performance improves with increasing
$N_{p}$, as expected from standard CS theory, and has similar dependence
on $\alpha$ as the oracle estimator, i.e., very good performance
is achieved in the large $\alpha$ regime even with small $N_{p}$.
Figure \ref{fig:MSE-vs-s} also depicts the $\mathsf{MSE}_{\text{tot}}$
expression of (\ref{eq:MSE_bound}) with optimal $s$. It can be seen
that it can serve as a reasonable lower bound for the CS-based estimator
performance for large $\alpha$, whereas this is not the case for
small $\alpha$, since, as discussed previously, the bound is not
tight in this regime. 

\section{Conclusion}

This paper considered the mean squared error performance of the oracle
estimator as an attempt to identify the limits of practical CS-based
techniques towards reducing downlink training signaling overhead in
CRAN. Using tools from stochastic geometry, an upper bound for the
oracle estimator performance was obtained in closed form, clearly
demonstrating the effects of design parameters, e.g., training sequence
length, and operational conditions, e.g., path loss exponent. It was
shown that good estimation performance can be expected with significant
training overhead reduction only under sufficiently large path loss
exponent.

\section*{Acknowledgment}

This work has been performed in the framework of the Horizon 2020
project ONE5G (ICT-760809) receiving funds from the European Union.
The authors would like to acknowledge the contributions of their colleagues
in the project, although the views expressed in this contribution
are those of the authors and do not necessarily represent the project.
The work of G. Wunder was also supported by DFG grants WU 598/7-1
and WU 598/8-1 (DFG Priority Program on Compressed Sensing).

\appendices{}

\section{Proof of Proposition \ref{thm: covariance of interference}}

Let $g_{s}>0$ denote the received power form the RRH with the $s$-th
largest modulus channel gain. Expressing $\|\mathbf{h}_{\bar{\mathcal{S}}}\|^{2}$
as 
\[
\|\mathbf{h}_{\bar{\mathcal{S}}}\|^{2}=\sum_{x\in\Phi}\mathbb{I}(|c_{x}|^{2}\|x\|^{-\alpha}<g_{s})|c_{x}|^{2}\|x\|^{-\alpha},
\]

\noindent where $\mathbb{I}(\cdot)$ is the indicator ($0-1$) function,
it follows that 
\begin{align*}
\mathbb{E}(\|\mathbf{h}_{\bar{\mathcal{S}}}\|^{2}) & \overset{(a)}{=}\lambda\int_{\mathbb{R}^{2}}|x|^{-\alpha}\mathbb{E}\left(\mathbb{I}(|c_{x}|^{2}|x|^{-\alpha}<g_{s})|c|^{2}\right)dx\\
 & \overset{(b)}{=}\lambda2\pi\int_{0}^{\infty}r^{1-\alpha}\mathbb{E}\left(\mathbb{I}(|c|^{2}r^{-\alpha}<g_{s})|c|^{2}\right)dr\\
 & \overset{(c)}{=}\lambda2\pi\mathbb{E}\left(|c|^{2}\int_{(|c|^{2}/g_{s})^{1/\alpha}}^{\infty}r^{1-\alpha}dr\right)\\
 & =\frac{2\pi\lambda\mathbb{E}\left(|c|^{4/\alpha}\right)}{\alpha-2}\mathbb{E}\left(g_{s}^{1-2/\alpha}\right)
\end{align*}

\noindent where ($a$) is an application of Cambell's Theorem \cite[Theorem A. 2]{Haenggi Ganti book},
($b$) follows by switching to polar coordinates for the integration
and dropping the explicit dependence of $c$ on $x$ (by definition
$c_{x}$ is independent of $x$), and ($c)$ follows by switching
the order of integration and expectation (Fubini's theorem).

In order to obtain $\mathbb{E}(g_{s}^{1-2/\alpha})$, the probability
distribution function (pdf) of $g_{s}$ is pursued next. To this end,
note that $g_{s}$ represents the $s$-th largest element of the one-dimensional
point process $\Pi\triangleq\{g_{x}=|c_{x}|^{2}\|x\|^{-\alpha}\}_{x\in\Phi}$
obtained by a transformation of the points of $\Phi$. It can be shown
that $\Pi$ is an \emph{inhomogeneous} PPP with density function $\lambda_{\Pi}(g)=\frac{2\pi}{\alpha}\lambda\mathbb{E}\left(|c_{x}|^{4/\alpha}\right)g^{-(1+2/\alpha)},g\geq0$
\cite[Theorem 4.1]{Mukherjee book}. This, in turn, means that the
number $N_{>\delta}$ of points in $\Pi$ that are greater than a
value $\delta\geq0$ is a Poisson random variable of mean
\begin{align*}
\mathbb{E}(N_{>\delta}) & =\int_{\delta}^{\infty}\lambda_{\Pi}(g)dg\\
 & =\pi\lambda\mathbb{E}(|c_{x}|^{4/\alpha})\delta^{-2/\alpha}.
\end{align*}
Following the approach of \cite{Haenggi distances}, the cumulative
distribution function (cdf) of $g_{s}$ equals
\begin{align*}
\mathbb{P}(g_{s}<\delta) & =\mathbb{P}(N_{>\delta}<s)=\sum_{k=0}^{s-1}\mathbb{P}(N_{>\delta}=k)\\
 & =\sum_{k=0}^{s-1}\frac{(\pi\lambda\mathbb{E}(|c_{x}|^{4/\alpha})\delta^{-2/\alpha})^{k}}{k!}e^{-\pi\lambda\mathbb{E}(|c_{x}|^{4/\alpha})\delta^{-2/\alpha}},
\end{align*}

\noindent and differentiation of the cdf w.r.t. $\delta$ gives the
probability distribution function (pdf) of $g_{s}$ as
\[
f_{g_{s}}(\delta)=\frac{2\left(\lambda\pi\mathbb{E}(|c_{x}|^{4/\alpha})\right)^{s}e^{-\lambda\pi\mathbb{E}(|c_{x}|^{4/\alpha})\delta^{-2/\alpha}}}{\delta^{\frac{2s}{\alpha}+1}\alpha(s-1)!},\delta\geq0.
\]

\noindent $\mathbb{E}(g_{s}^{1-2/\alpha})$ can now be computed as
$\int_{0}^{\infty}f_{g_{s}}(\delta)\delta^{1-2/\alpha}d\delta$, which
has a closed form expression for $s>\frac{\alpha}{2}-1$, finally
leading to (\ref{eq:covariance of interference}).

\end{document}